\documentclass[12pt,preprint]{aastex}
\usepackage{graphics}

\def\kpc {\rm{Kpc}}
\def\and  {\it {et al.} \rm}
\def\rmd {\rm d}

\def\fun#1#2{\lower3.6pt\vbox{\baselineskip0pt\lineskip.9pt
\ialign{$\mathsurround=0pt#1\hfill##\hfil$\crcr#2\crcr\sim\crcr}}}

\def\spose#1{\hbox to 0pt{#1\hss}}
\def\simlt{\mathrel{\spose{\lower 3pt\hbox{$\mathchar"218$}}
     \raise 2.0pt\hbox{$\mathchar"13C$}}}
\def\simgt{\mathrel{\spose{\lower 3pt\hbox{$\mathchar"218$}}
     \raise 2.0pt\hbox{$\mathchar"13E$}}}
\def\beq{\begin{equation}}
\def\eeq{\end{equation}}
\def\bce{\begin{center}}
\def\ece{\end{center}}
\def\bea{\begin{eqnarray}}
\def\eea{\end{eqnarray}}
\def\ben{\begin{enumerate}}
\def\een{\end{enumerate}}

\def\brr{\begin{array}}
\def\err{\end{array}}

\begin{document}

\title{Statistical Properties of Galactic Starlight Polarization}

\author{
Pablo Fosalba\altaffilmark{1,4}, Alex Lazarian\altaffilmark{2}, 
Simon Prunet\altaffilmark{3}, Jan A. Tauber\altaffilmark{4}}

\altaffiltext{1}{Institut d'Astrophysique de Paris, 98bis Boulevard
Arago, F-75014 Paris, France; E-mail: fosalba@iap.fr} 
\altaffiltext{2}{Department of Astronomy, University of Wisconsin, 
Madison, USA; E-mail: lazarian@astro.wisc.edu} 
\altaffiltext{3}{Canadian Institute for Theoretical Astrophysics, 
McLennan Labs 60 St George Street, Toronto ON M5S 3H8, Canada; 
E-mail: prunet@cita.utoronto.ca}
\altaffiltext{4}{Astrophysics Division, SSD of ESA-ESTEC, P.O. Box 299, 
NL-2200 AG Noordwijk, The Netherlands; E-mail: jtauber@astro.estec.esa.nl}


\begin{abstract}

We present a statistical analysis 
of Galactic interstellar polarization from the largest compilation 
available of starlight data. 
The data comprises $\sim 9300$ stars 
of which we have selected 
$\sim 5500$ for our analysis.  
We find a nearly linear growth of mean polarization degree with
extinction. 
The amplitude of this correlation shows 
that interstellar grains are not fully aligned with the Galactic magnetic field,
which can be interpreted as the effect of a large random component of the field.
In agreement with earlier studies of more limited scope,
we estimate the ratio of the uniform to the random plane-of-the-sky
components of the magnetic field to be ${\bf B_u/B_r} \approx 0.8$. 
Moreover, a clear correlation exists between polarization degree 
and polarization angle
what provides evidence that the magnetic field geometry
follows Galactic structures on large-scales.

The angular power spectrum $C_{\ell}$ of the starlight polarization degree
for Galactic plane data ($|{\rm b}| < 10^{\circ}$) is
consistent with a power-law, $C_{\ell} \propto \ell^{-1.5}$ 
(where $\ell \approx 180^{\circ}/\theta$ is the multipole order),
for all angular scales $\theta \simgt 10^{\prime}$.  
An investigation of 
sparse and inhomogeneous sampling of the data shows that
the starlight data analyzed traces an underlying polarized continuum
that has the same power spectrum slope, $C_{\ell} \propto \ell^{-1.5}$.
Our findings suggest that starlight data can be safely used for the modeling
of Galactic polarized continuum emission at other wavelengths.

\end{abstract}

\keywords{dust, extinction ---
	  Galaxy: stellar content, structure ---
	  methods: statistical ---
	  polarization} 


\clearpage

\section{Introduction}

The large scale emission from the Milky Way at radio, mm-wave and far-infrared 
wavelengths
is known to be polarized (see e.g, de Oliveira-Costa \& Tegmark 1999).
The underlying common cause is the
Galactic magnetic field, though the particular polarization mechanism is
wavelength-dependent. It ranges from the synchrotron emission process which is the driving 
mechanism at long wavelengths to the absorption/emission properties of aligned
dust grains which are expected to account for the short-wavelength emission. 
Therefore the measurement of the polarized
Galactic emission should yield valuable information on our Galaxy's magnetic
field (see e.g, Zweibel \& Heiles 1997, Hildebrand et al 2000, 
Heitsch et al 2001).
In particular, 
starlight polarization vectors trace the plane-of-the-sky
projection of the Galactic magnetic field (Zweibel \& Heiles 1997) and
measurements of polarization for stars of different distances reveals
the 3D distribution of magnetic field orientations averaged along the line of sight.

However currently the only available large scale maps of polarized Galactic emission
are at low (radio) frequencies 
(i.e, $\nu \le 2.7$ GHz, see Tucci et al 2000, 
for a description of available data), 
where the Galactic signature is
significantly distorted by (relatively) local Faraday rotation effects. 
At far-infrared and mm wavelengths, measurements have been made of a 
few very small 
regions, largely dense dark clouds, mainly concentrated in the Galactic plane 
(Hildebrand et al 1999, Novak et al 2000; see also Hildebrand et al 2000,
Heitsch et al 2001, for recent reviews)
but they reflect only rather local distortions of the large-scale magnetic field.
Therefore, within this wavelength range, the only large scale view has been obtained
by observation of and analogy with external spiral galaxies similar to our
own (Zweibel \& Heiles 1997). 

The next generation of Cosmic Microwave background (CMB) missions
(i.e. MAP - \\
{\tt http://map.gsfc.nasa.gov/} - and Planck - 
{\tt http://astro.estec.esa.nl/Planck}) will fill this gap by providing polarization
measurements of the whole sky at mm wavelengths. For these missions, the
polarized Galactic dust emission is primarily a nuisance 
which has to be removed from the underlying
cosmic signal (see Prunet \& Lazarian 1999, Draine \& Lazarian 1999,
Lazarian 2001), 
but in so doing they will also provide for the first time the means to
map the Galactic magnetic field. Planck in particular covers a range of 
wavelengths (1 cm to 300 $\mu$m)
in which different polarization mechanisms are dominant, which will provide
powerful means to uncover the common underlying magnetic field distribution.

At optical wavelengths, 
there do exist many measurements of starlight polarization \\
(Appenzaller 1974, Schroeder 1976, Mathewson et al 1978,
Markkanen 1979, Krautter 1980, Korhonen \& Reiz 1986,  
Bel et al 1993, Leroy 1993, Berdyugin et al 1995, 
Reiz \& Franco 1998). 
Observed starlight polarization is
believed to be caused by selective absorption by 
magnetically 
aligned interstellar dust grains along the line of sight; 
because these measurements are limited by dust extinction, they
give us a view of the behavior of the magnetic field only in a local bubble around
us. Furthermore, the pencil-beam nature of these observations means that the 
view afforded is punctual, with a sampling that is both inhomogeneous and sparse.
However, analysis of compilations of such measurements 
(see Heiles 2000 and references therein) 
imply that they do contain information about both the large-scale and the random 
components of the Galactic field in the vicinity of the Sun.
Therefore it is useful to extract as much information as possible from this data.

In this paper we analyze the most complete compilation to date 
of optical polarization observations. This analysis will allow us to 
extract basic information on the
large scale statistical properties of the polarization field in the visible.
We do this by studying the correlations between stellar parameters and
computing the angular power spectrum 
(a convenient statistical estimator for analyzing 2D full-sky maps)
of the optical polarization degree from Milky Way stars; in so doing
we investigate the systematic effects introduced by the sparse and inhomogeneous
sampling which is intrinsic to starlight measurements.

The outline of the paper is as follows:
\S\ref{sec:data} presents the starlight data analyzed.
We describe the 
distribution of the sources in \S\ref{sec:distri} and study their correlations
as a function of distance and Galactic coordinate in \S\ref{sec:correla}.
We compute the angular power spectrum of starlight polarization degree 
in \S\ref{sec:ps}. We finish with a discussion
of our results in \S\ref{sec:disc}.

\section{Data}
\label{sec:data}

The starlight polarization data
used in this analysis is taken from the compilation by Heiles 2000.
This compilation includes data from 9286 sources taken from
a dozen of catalogs, combining multiple observations, providing accurate
positions and reliable estimates for extinction and distance of stars.

From this catalog, we have selected a subsample of 
5513 stars ($60 \%$ of the data) based on the following criteria:

\begin{itemize}
\item{the degree and angle of polarization are given}
\item{small absolute error in the polarization degree ($< 0.25 \%$)
\footnote{This criterium in principle allows for the 
inclusion of sources with low
statistical significance in the polarization degree. However, the
mean {\em relative} error of such sources is comparable to that of the
entire sample of 5513 stars. 
This reflects that large {\em relative} error bars are intrinsic
to the polarization degree measurements available and
supports the criterium of leaving out sources with large 
{\em absolute} (rather than relative) error.}}

\item{a (positive) extinction is given}
\end{itemize}

Note that the most constraining requirement is the last one
(extinction): if this requirement were not used,
the subsample would include 8280 stars (90 $\%$ of the whole compilation).
However we need to include the degree of extinction in the visible as
it strongly correlates with the polarization degree (see
\S\ref{sec:correla} below) and it is a basic ingredient to model
dust polarization at other wavelengths
(Hildebrand \& Dragovan 1995). We shall present this analysis in a forthcoming
paper (Fosalba et al 2001).
All the stars in the Heiles compilation fulfilling the above requirements
also have quoted distance (with an estimated 20 $\%$ 
error for most of the sources,
see Heiles 2000).


\section{Distribution of Sources}
\label{sec:distri}

\begin{figure*}[t!]
\begin{center}
\includegraphics[width=12.0cm]{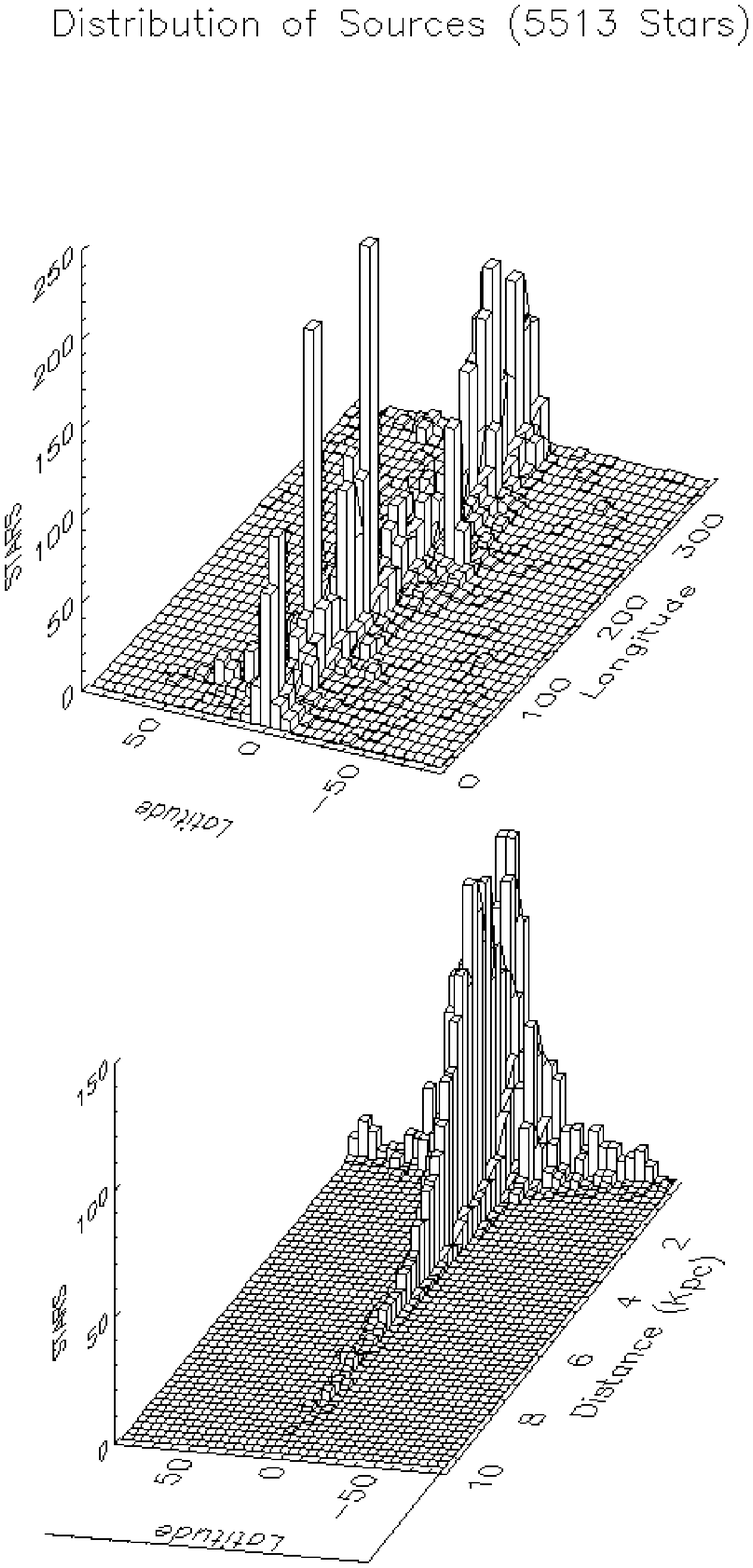}
\end{center}
\caption{\label{hist_star_lb_db}
Distribution of starlight polarization data in Galactic
longitude and latitude bins (top panel) and distance and latitude bins
(bottom panel) for the subsample of 5513 stars analyzed.}
\end{figure*}

Fig \ref{hist_star_lb_db} shows the distribution of sources in our subsample
for data binned in Galactic coordinates (top panel) as well as
in distance and latitude (bottom panel).
As shown in the latter,
all high latitude ($|{\rm b}| > 10^{\circ}$) stars are nearby (d $<$ 1 \kpc).
Within the Galactic plane one can find relatively distant stars, though the
vast majority are within 2 \kpc. Thus, this is clearly a local sample.


This is also seen in 
Table \ref{tab:sample}, which summarizes the mean stellar parameters 
(i.e, polarization degree P($\%$) 
and extinction as measured by the color excess E(B-V) )
in the subsample
as a function of latitude and distance.
Note that in Table \ref{tab:sample},
high latitude (low latitude) means 
$|{\rm b}| > 10^{\circ}$ ($|{\rm b}| < 10^{\circ}$)
and nearby (distant) denotes d $<$ 1 \kpc ~(d $>$ 1 \kpc).
The quantities between brackets denote amount $\%$ of all stars in the sample.
Low-latitude stars have large values of the polarization degree 
P($\%$) $\approx 1.7$, and extinction 
E(B-V) $\approx 0.5$, while high-latitude sources exhibit
significantly lower values, P($\%$) $\approx 0.5$, E(B-V) $\approx 0.15$.  
Polarization vectors (defined with respect to galactic coordinates) 
are typically oriented along the galactic plane 
($\theta_p \approx 90^{\circ}$) although a more detailed analysis
reveals a rich spatial distribution 
(see Fig \ref{polAngle_5513} and \S\ref{sec:correla} below).



\begin{figure*}[t!]
\begin{center}
\includegraphics[width=10.0cm]{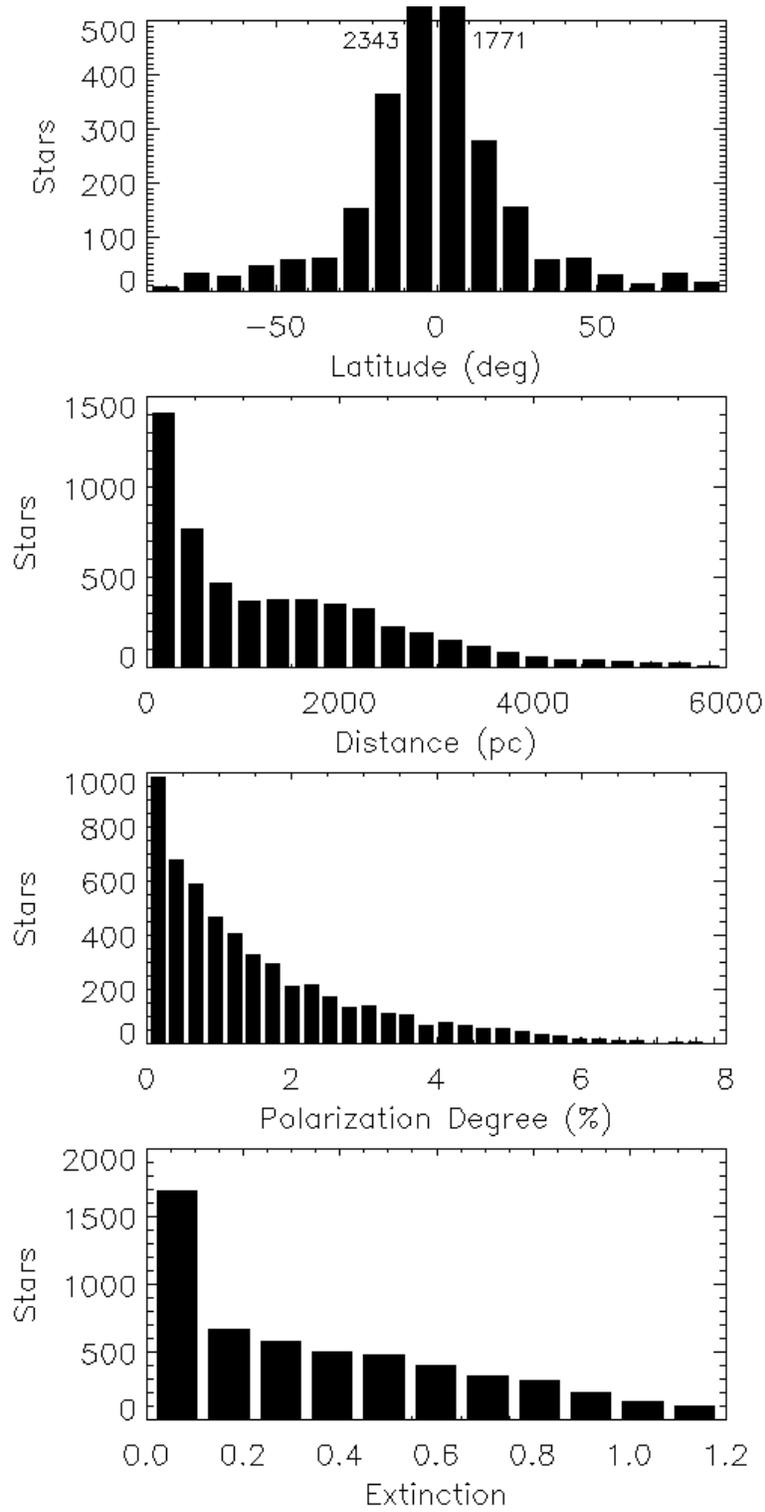}
\end{center}
\caption{\label{hist_star_bdpe}
Distribution of sources with Galactic latitude, distance, polarization
degree and extinction. In the top panel, the number of stars in the two central
bins are displayed beside them as they are far above the plotted range.}
\end{figure*}

Histograms displaying the distribution of sources
with Galactic latitude b, distance d, polarization degree P($\%$) 
and extinction E(B-V) bins, are given in Fig \ref{hist_star_bdpe}.


\section{Correlations between Stellar Parameters}
\label{sec:correla}

\begin{figure*}[t!]
\begin{center}
\includegraphics[width=18.0cm]{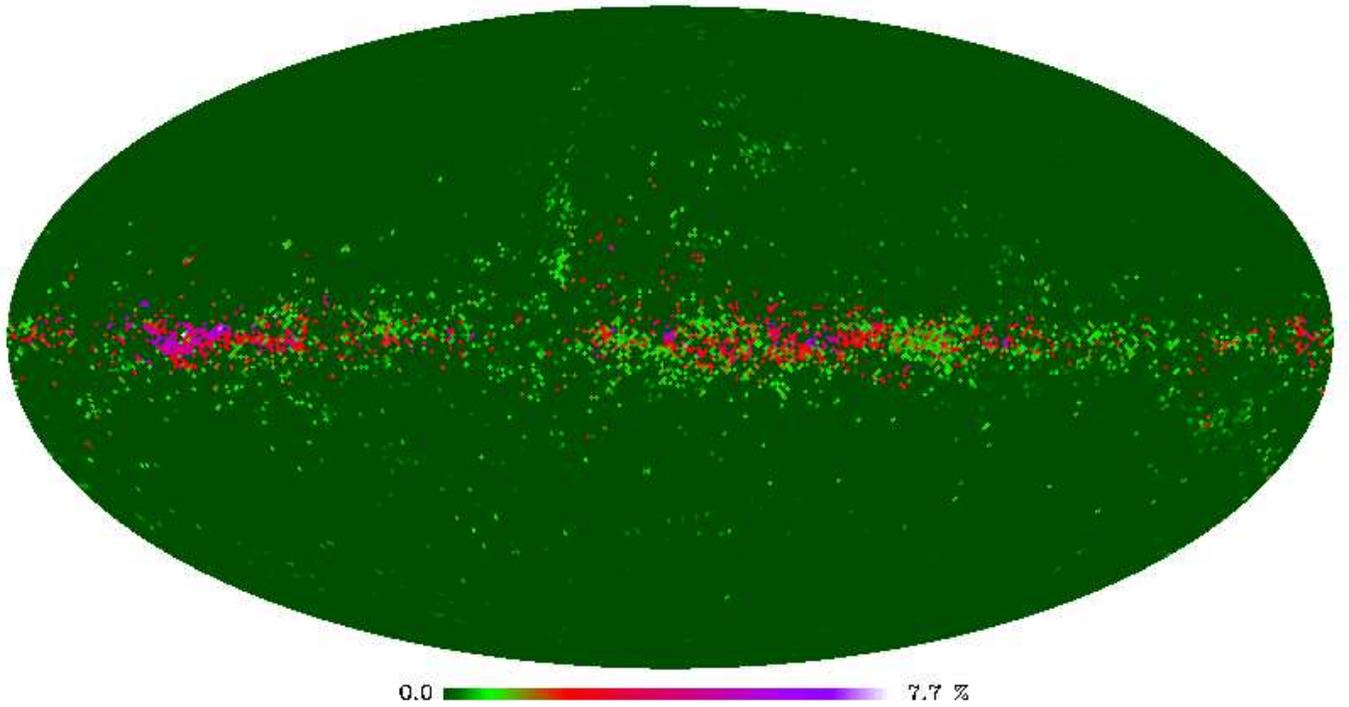}
\end{center}
\caption{\label{polDegree_5513}
Starlight Polarization Degree for the subsample of 5513 stars
in Galactic coordinates. The map is shown in HEALPix tessellation 
with a pixel resolution of $1^{\circ}$ for convenience. Nearby stars (light-green 
sources) lie mainly at high Galactic latitudes while distant stars (red-purple sources)
are found in the Galactic plane.}
\end{figure*}

\begin{figure*}[t!]
\begin{center}
\includegraphics[width=18.0cm]{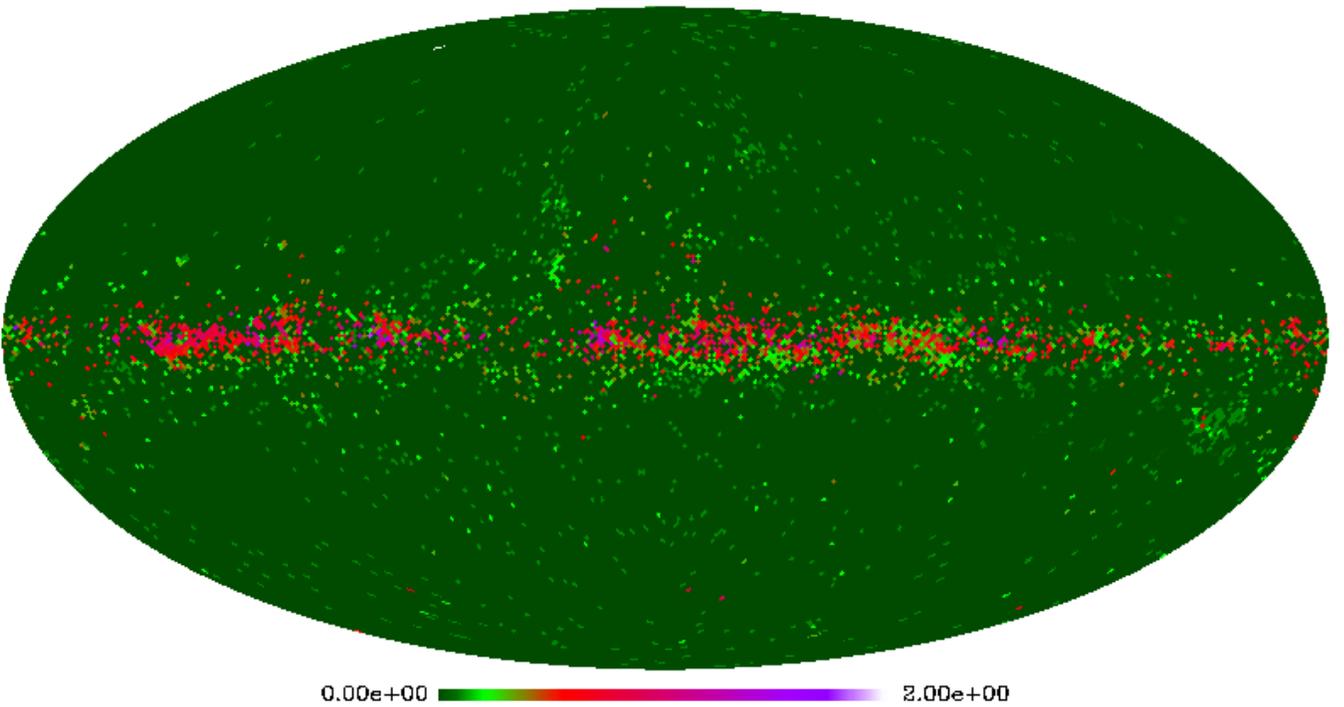}
\end{center}
\caption{\label{ebv_5513}
Same as Fig \ref{polDegree_5513} but for Starlight Extinction, E(B-V).
Sources with extinction E(B-V) $> 2$ are not shown for clarity.
A strong spatial correlation with the polarization degree distribution, 
Fig \ref{polDegree_5513}, is observed.}
\end{figure*}

Light emitted from stars is assumed to be unpolarized; 
the observed polarization from starlight is believed to result 
from extinction by interstellar dust grains along the line of sight 
to the observer.  Thus, for a homogeneous 
distribution of intervening dust, the larger
the path-length starlight travels to reach the observer, 
the larger the polarization degree
and extinction are expected to be. 
According to this simple picture, Galactic regions with low
polarization degree (and extinction) trace nearby stars, 
while regions with high values of these parameters
correspond to distant stars.
This is actually observed in the sample of Starlight data, as shown 
in Figs \ref{polDegree_5513} \& \ref{ebv_5513}, 
where nearby stars (light green sources) are mainly at high Galactic latitudes, 
while distant stars (red-purple sources) are found in the Galactic plane (see 
also Table \ref{tab:sample} for mean values of the stellar parameters).
In particular, 
the spatial distribution of both parameters in Galactic coordinates
is expected to be highly correlated and this is clearly observed in
Figs \ref{polDegree_5513} \& \ref{ebv_5513}.

We describe in detail below how the stellar parameters which describe
the sources in the subsample correlate with one another and what information 
can be extracted from the behavior found.

\subsection{Behavior with Distance}
\label{sec:dist}

\begin{figure*}[t!]
\begin{center}
\includegraphics[width=14.0cm]{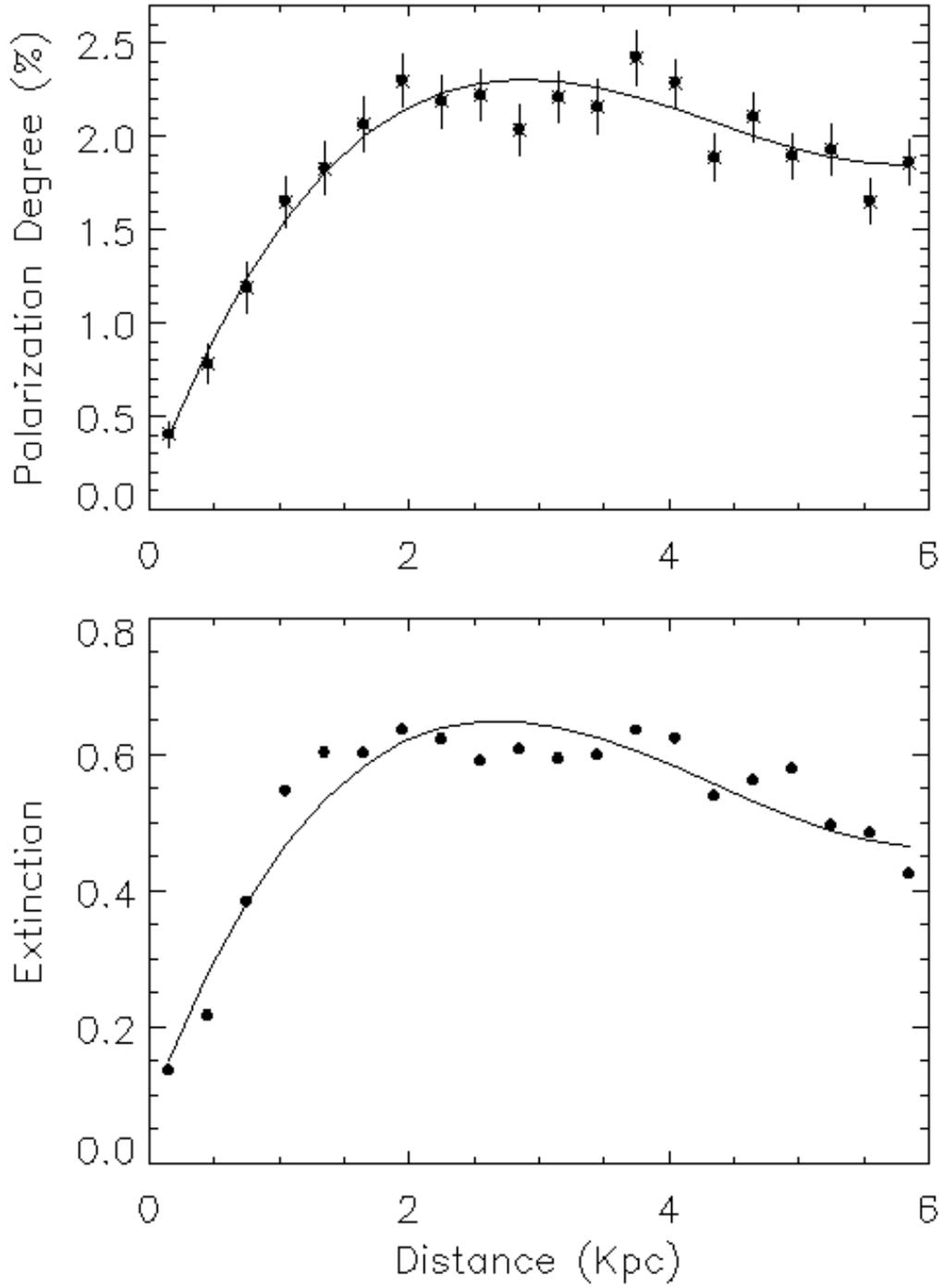}
\end{center}
\caption{\label{disbins_poldeg_ebv}
Polarization degree with (quoted) error bars (top panel)
and extinction (bottom panel) in linear distance bins.
Solid lines show best fit curves to third order polynomials, 
Eqs(\ref{eq:poldist}) \& (\ref{eq:ebvdist}).}
\end{figure*}

The distribution of the polarization degree and extinction as a function of distance
(average quantities in linear distance bins)
for the sources considered are shown in Fig \ref{disbins_poldeg_ebv}. 
It is seen that both stellar parameters grow linearly (to a good approximation)
with distance up to d $\approx 2$ \kpc.
Beyond $\sim$2 \kpc, stars have
roughly constant values for both quantities, i.e,
P($\%$) $\approx$ 2, E(B-V) $\approx$ 0.6 . 
The overall behavior of P($\%$) and E(B-V) with distance (in \kpc) can be best-fitted 
by third-order polynomials up to d $\approx 6$ \kpc 
(see Fig \ref{disbins_poldeg_ebv}):
\beq
{\rm P}(\%) \approx 0.13 + 1.81 \rmd - 0.47 \rmd^2 + 0.036 \rmd^3 ~ ,
\label{eq:poldist}
\eeq
\beq
{\rm E(B-V)} \approx 0.08 + 0.5 \rmd - 0.135 \rmd^2 + 0.0104 \rmd^3 ~.
\label{eq:ebvdist}
\eeq

\begin{figure*}[t!]
\begin{center}
\includegraphics[width=14.0cm]{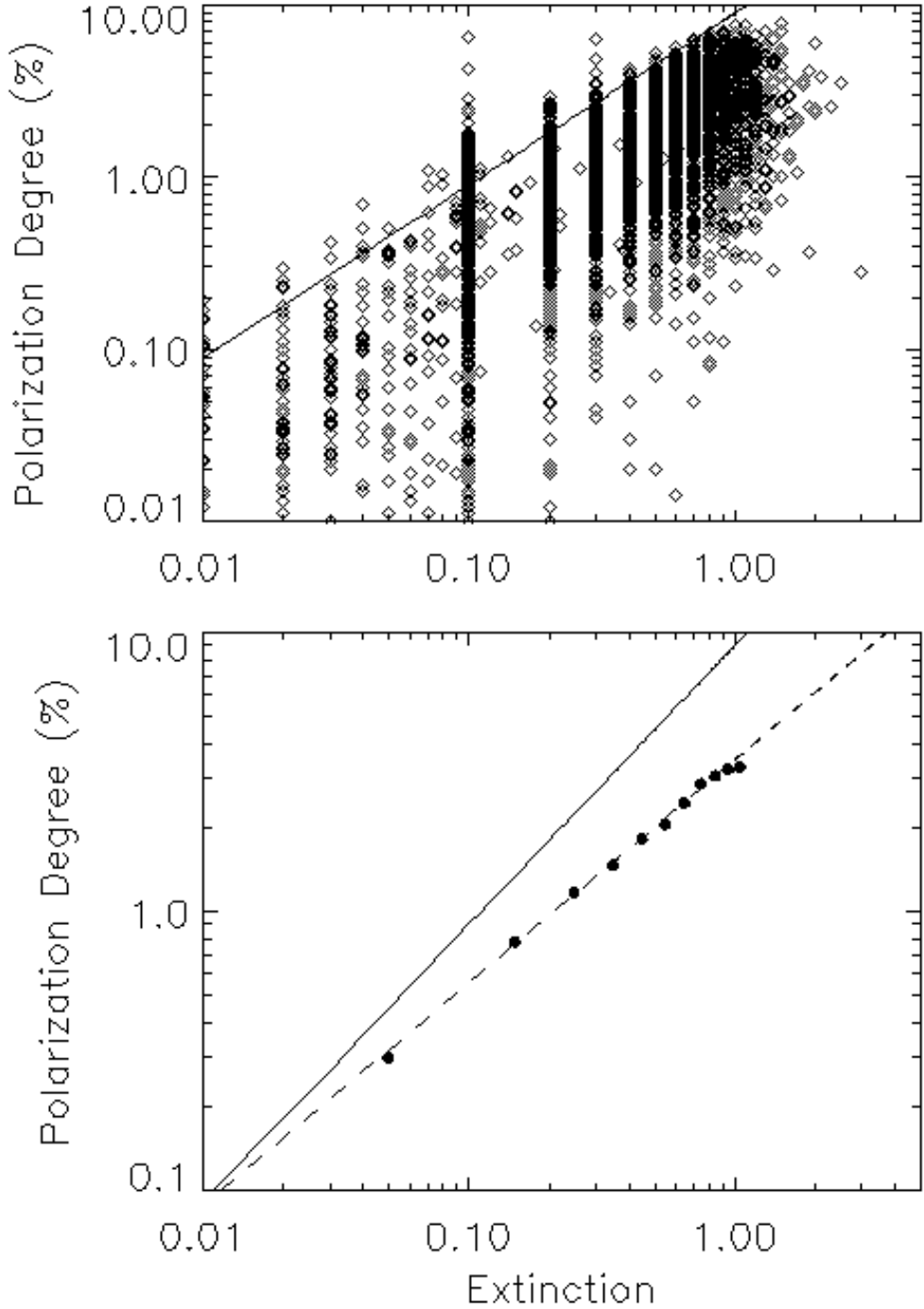}
\end{center}
\caption[junk]{\label{ebvbins_poldeg}
Correlation between polarization degree P($\%$), and extinction E(B-V).
Upper panel shows all individual sources
while lower panel displays data averaged in extinction bins.
Solid line shows the theoretical upper limit, P($\%$) = 9 E(B-V), for
completely aligned grains by external (regular) magnetic fields.
Dashed line in lower panel shows P($\%$) = 0.39 E(B-V)$^{0.8}$, 
which is a good fit to the data up to E(B-V) $\approx 1$.}
\end{figure*}

The similar behavior of both stellar parameters with distance already suggests a
simple linear correlation (for data averaged in distance bins)
between polarization degree and extinction. 
This is the case indeed
as shown in Fig \ref{ebvbins_poldeg}.
The observed roughly linear correlation for individual sources 
is in agreement with measurements at 2.2 $\mu$m (Jones 1989)
\& 100 $\mu$m (Hildebrand et al 1995).

However the observed mean correlation amplitude for data averaged in
extinction bins is much smaller than  
what is expected from interstellar dust grains completely aligned under a purely
regular (no random component) external magnetic field 
P($\%$) $\approx$ 9 E(B-V). This is also observed in the K-filter at
2.2 $\mu$m  (Jones 1989). What is more,
we find a slight deviation from the simple linear correlation, 
\beq
{\rm P(\%)} \approx 3.5 ~{\rm E(B-V)}^{0.8} ,
\label{eq:polext}
\eeq 
as displayed in the lower panel of Fig \ref{ebvbins_poldeg}.
In fact, this deviation from the linear correlation shows a similar dependence
with extinction for the K-band observations at
2.2 $\mu$m, i.e, \\
${\rm P_k} \approx$  0.83 E(B-V)$^{0.75}$ (Jones 1989).  
Eq(\ref{eq:polext}) implies that the ratio of the observed 
to the optimal polarization degree,
\beq
{\rm {P_{Obs} \over P_{Max}}} \approx 0.39 ~{\rm {E(B-V)}}^{-0.2}
\label{eq:polratio}
\eeq 
This is, at least, a factor of 2 larger than the same ratio found for the
K-filter at 2.2 $\mu$m, ${\rm P_{Obs}/P_{Max}} \approx 0.19$ 
(Jones 1989). 
Next we study 
the dependence of the stellar parameters with Galactic coordinate.


\subsection{Behavior with Galactic Longitude}
\label{sec:londep}

\begin{figure*}[t!]
\begin{center}
\includegraphics[width=15.0cm]{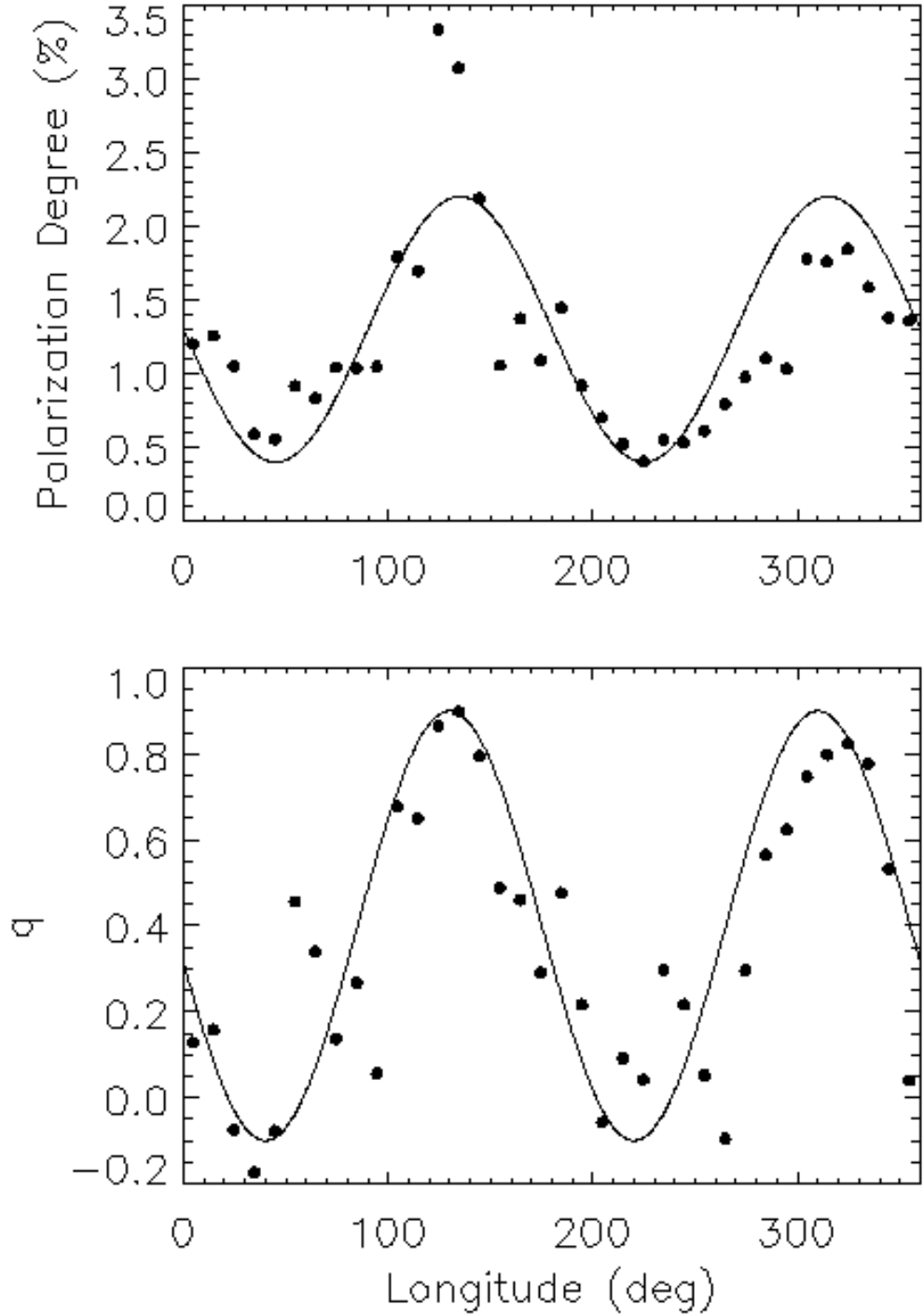}
\end{center}
\caption{\label{lonbins_poldeg_stokesQ}
Starlight Polarization Degree (top panel) and
q parameter (bottom panel) for data averaged in 10$^{\circ}$ 
longitude bins.
The solid line shows a best fit to a sinusoidal dependence, 
Eqs(\ref{eq:polon}) \& (\ref{eq:qlon}).}
\end{figure*}

In order to study the dependence of stellar parameters with Galactic
longitude we have averaged the data in 10$^{\circ}$ longitude bins. 
As shown in Fig \ref{lonbins_poldeg_stokesQ}, both
the polarization degree and the q parameter, defined
as\footnote{Our definition of the q parameter corresponds to
the so-called {\em fractional polarization} and it is related to the
Stokes Q parameter, 
$Q = I q$, where I is the polarized intensity} 
$q = \cos 2 (\theta_p - 90^{\circ})$ ($\theta_p$ is the polarization angle),
exhibit, on average, 
a sinusoidal-like dependence with longitude with a 180$^{\circ}$
periodicity, well-fitted by the expressions,
\beq
{\rm P}(\%) \approx 1.3 + 0.9 \sin (2 \,l + 180^{\circ}) ~,
\label{eq:polon}
\eeq
\beq
{\rm q} \approx 0.4 + 0.5 \sin (2 \,l + 190^{\circ}) ~ . 
\label{eq:qlon}
\eeq

\begin{figure*}[t!]
\begin{center}
\includegraphics[width=18.0cm]{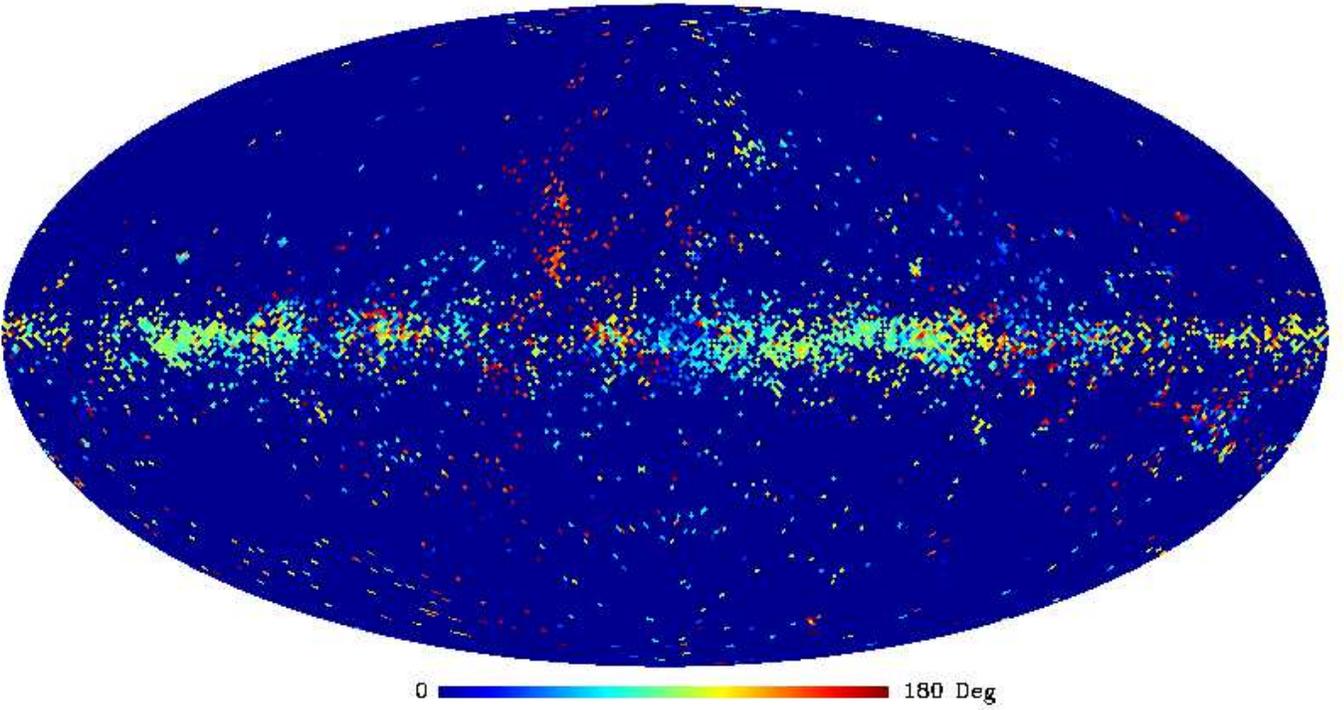}
\end{center}
\caption{\label{polAngle_5513}
Same as Fig \ref{polDegree_5513} but for the Starlight Polarization Angle.
Note that, on large-scales, the polarization vectors are mostly aligned 
with the Galactic plane structure (light-green, light-blue and yellow sources)
except for the spherical shell of Loop 1, where polarization vectors
are roughly perpendicular to the Galactic plane (red sources).}
\end{figure*}

For Galactic longitudes $l \approx 50^{\circ}$ and
$l \approx 230^{\circ}$  
we find minimum values of q, i.e, the stellar polarization vectors are orthogonal 
to the Galactic plane, $\theta_p \approx 0^{\circ}, 180^{\circ}$. 
Note that, at the Galactic plane, these directions approximately intersect
the Cygnus-Orion spiral arm which suggests that, on average, 
polarization vectors   
do there not align with this Galactic structure.
Moreover, we also find minimum values of the polarization degree 
(or extinction as they are nearly linearly correlated) 
for these Galactic longitudes.
Approximately along these directions (as one moves away from the Galactic plane) 
one finds the edge of a supernova remnant, the spherical shell of Loop I (see
red sources in Fig \ref{polAngle_5513}). 
Thus, a possible explanation for the values of the stellar 
parameters along these directions is that exploding supernovae
in the Scorpius/Ophiuchus star cluster centered at $l \approx 0^{\circ}$,
b $\approx 20^{\circ}$ (see Zweibel \& Heiles 1997) 
could cause polarization vectors
to be strongly aligned with the Galactic structure left by the 
supernova remnant.

On the other hand, maximum values of the polarization degree and q 
parameter are found at
$l \approx 140^{\circ}$ and $l \approx 320^{\circ}$, where the polarization vectors
of dust grains are parallel to the Galactic disk structure,
$\theta_p \approx 90^{\circ}$ (see light-green sources in Fig \ref{polAngle_5513}).

We note that the results shown in the lower panel of
Fig \ref{lonbins_poldeg_stokesQ} for the longitude dependence of
the q parameter are in good agreement
with the analysis presented in Whittet (1992)
for about 1000 nearby Galactic plane stars
(d $< 0.6 ~ \kpc$, $|{\rm b}| < 3^{\circ}$).


\subsection{Behavior with Galactic Latitude}
\label{sec:latdep}

As discussed in \S\ref{sec:distri}, 
most of the sources in our subsample 
are in the Galactic disk (75 $\%$ of the stars are found 
at $|{\rm b}| < 10^{\circ}$, see Table \ref{tab:sample}). 
However, there is a statistically significant fraction of the sources 
at high Galactic latitudes (25 $\%$ of sources at  $|{\rm b}| > 10^{\circ}$) 
which allow us to investigate  
the mean variation of the correlations between stellar parameters 
as a function of latitude.
For this purpose we have averaged the data in $10^{\circ}$ (linear) latitude bins. 

\begin{figure*}[t!]
\begin{center}
\includegraphics[width=15.0cm]{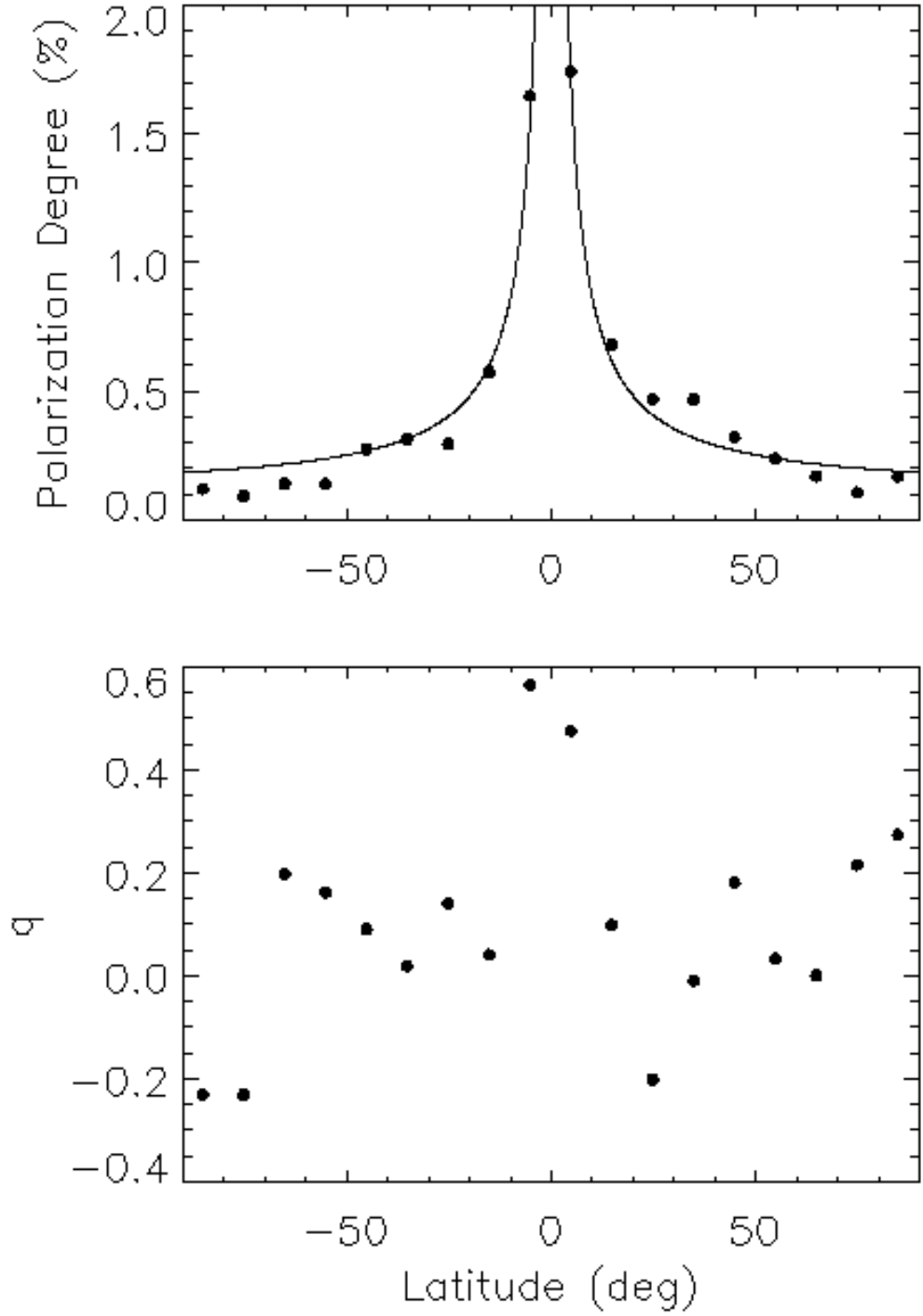}
\end{center}
\caption{\label{latbins_poldeg_stokesQ}
Starlight Polarization Degree (top panel) and
q parameter (bottom panel) for data averaged in 10$^{\circ}$ 
latitude bins. The solid line in the upper panel shows a co-secant law 
distribution, Eq(\ref{eq:polat}).}
\end{figure*}

We find that the polarization degree shows 
a strong dependence with latitude
as shown in the upper panel of Fig \ref{latbins_poldeg_stokesQ}.
Indeed, the behavior can be well described by a co-secant law:
\beq
{\rm P}(\%) \approx 0.1 + 0.0067 \csc (0.05 ~|{\rm b}|) .
\label{eq:polat}
\eeq

The statistical significance in the measurement 
of the q parameter (i.e, the polarization angle) 
is low at high latitudes 
given the lack of available data and the large 
degree of dispersion in the polarization angle
\footnote{Only the two data points of the galactic plane bins 
seem to suggest that polarization angles follow galactic structures
($\theta_p \approx 90^{\circ}$), but no significant correlation is observed.} 
(as large as 35-55 $\%$, depending on latitude and distance; 
see also the large scatter in lower panel of Fig \ref{lonbins_poldeg_stokesQ}). 
Therefore it is hard 
to draw any conclusions about the 
correlation between the polarization degree and q
parameter on high Galactic latitudes from the current analysis.


\section{The Angular Power Spectrum of the Starlight Polarization Degree}
\label{sec:ps}

In this section, we analyze the starlight polarization data 
described in \S\ref{sec:data} 
to see whether and if so, to what extent, 
it traces from a statistical point of view
the {\em large-scale pattern of starlight polarization}. 
By this we mean
that ISM dust grains provide a discrete (and high-resolution)
picture of the diffuse polarizing ISM along a few lines-of-sight, from
which one aims at reconstructing (using in particular the power
spectrum analysis
presented below) the large-scale (low-resolution) 
statistical properties of the ISM polarization pattern.

In fact, the large-scale statistical properties of the
ISM polarization from absortion of starlight by dust grains might give
direct statistical information on the polarized diffuse emission by
dust: if the grains that extinct starlight and emit constitute the same grain
population, the power spectrum 
of starlight polarization degree is directly related to the power
spectrum of polarized emission from dust. 

Starlight polarization is caused by
aligned grains with sizes $10^{-4}>a>10^{-5}$~cm (Kim \& Martin 1995).
These are essentially the grains responsible in diffuse media for
polarized emission (see Prunet \& Lazarian 1999). 
Therefore if aligned grains
in diffuse medium have the same temperature the power spectrum 
of the starlight
polarization should be identical to the spectrum of the polarized
continuum from dust in the far infrared range (e.g, $100 \mu$ m).
The temperature difference between graphite and silicate grains
will not change this result as graphite grains are usually not
aligned (see Whittet 1992). In diffuse media, unlike in molecular
clouds, the grains are exposed to a similar radiation flux and
find themselves in a very similar physical environment. Therefore
we do not expect substantial variations of the grain temperature
and believe that the power spectrum we find will also represent the power
spectrum of the polarized dust emission. 

For this purpose we estimate the {\em angular} power spectrum (PS hereafter)
of the starlight polarization degree (see Fig \ref{polDegree_5513}).
The angular power spectrum is the convenient full-sky generalization of the 2D 
Fourier power spectrum, as the latter is only strictly valid 
in small (flat) patches of the sky. 
In particular, given a full-sky map of a scalar 
field {\rm S} (such as the starlight polarization degree) 
one can decompose it into
spherical harmonic basis, $Y_{\ell m}$, 
at any point ($\theta,\phi$) in the sky:
\beq
S(\theta,\phi) = \sum_{\ell m} a_{\ell m} Y_{\ell m} (\theta,\phi) \,.
\eeq
The spherical harmonic basis, $Y_{\ell m}$, is the 
generalization of the Fourier basis to the sphere 
(as opposed to the usual flat space description).  
The $a_{\ell m}$'s are the amplitudes 
of the decomposition of a scalar field on this basis and
they are assumed to be random Gaussian variables.
\footnote{The $a_{\ell m}$'s are taken to be 
a realization of a
random field whose amplitudes at a given point 
have a Gaussian distribution. In Fourier space, this is equivalent to
say that the statistical properties of the field are completely
determined by its power spectrum.}

The angular power spectrum is thus defined as a quadratic average of
a given $a_{\ell m}$ coefficient over different m-modes:
\beq
C_{\ell} = {1 \over {2 \ell +1}} \sum_{m=- \ell}^{m=\ell} | a_{\ell m}|^2 
\label{eq:cl}
\eeq
where $C_l$ estimates autocorrelations of the field at an angular scale 
$\theta \approx 180^{\circ}/ \ell$, being 
$\ell$ the so-called multipole order.
In a fully-sampled map, the two-point correlation function
$\xi(\theta)$ of the scalar field ${\rm S}$ 
is simply related to the angular power spectrum:
\beq
\xi (\theta) = <S({\bf q_1}) S({\bf q_2})> = 
\sum_{\ell} {\ell+1/2 \over {\ell(\ell +1)}} C_{\ell} P_{\ell} (\cos \theta)
\label{eq:xi} 
\eeq
where $\cos \theta = {\bf q_1} \cdot {\bf q_2}$ 
is the dot product of two unit vectors pointing 
to any pair of sky pixels, $P_{\ell}$ is the Legendre
polynomial of multipole order $\ell$ and $<S>$ denotes ensemble
average of the field $S$.

Although the angular power spectrum $C_{\ell}$ 
is a straightforward statistic to calculate, 
its computation is usually limited to fully-sampled maps. 
As such, it conveys information at all
angular scales down to the map resolution. 
Note that for the starlight data, a map of very large spatial resolution
is required as the information is essentially point-like.
Furthermore significant systematic effects are expected to be 
introduced by the 
very sparse and inhomogeneous (clustered) nature of this data sample.

\subsection{Power Spectrum Estimation of the Real Data}
\label{sec:pse}

We want to assess to what extent the discrete data available on
starlight polarization is able to trace an underlying large-scale 
polarized pattern in the visible.
This problem is a particular case of a more general one, say, 
what information about the (large-scale) continuum signal can be extracted
from sparse in-homogeneously distributed discrete data.

In order to address this issue, we shall
investigate how the latter effects 
affect the angular power spectrum estimation.
For this purpose, we have analyzed the Galactic plane data 
($|{\rm b}| < 10^{\circ}$),
which is the most densely sampled in the catalog and therefore it is expected
to yield the most reliable statistics.

\subsubsection{Rough Harmonic Analysis}
\label{sec:rough}

\begin{figure*}[t!]
\begin{center}
\includegraphics[width=16.0cm]{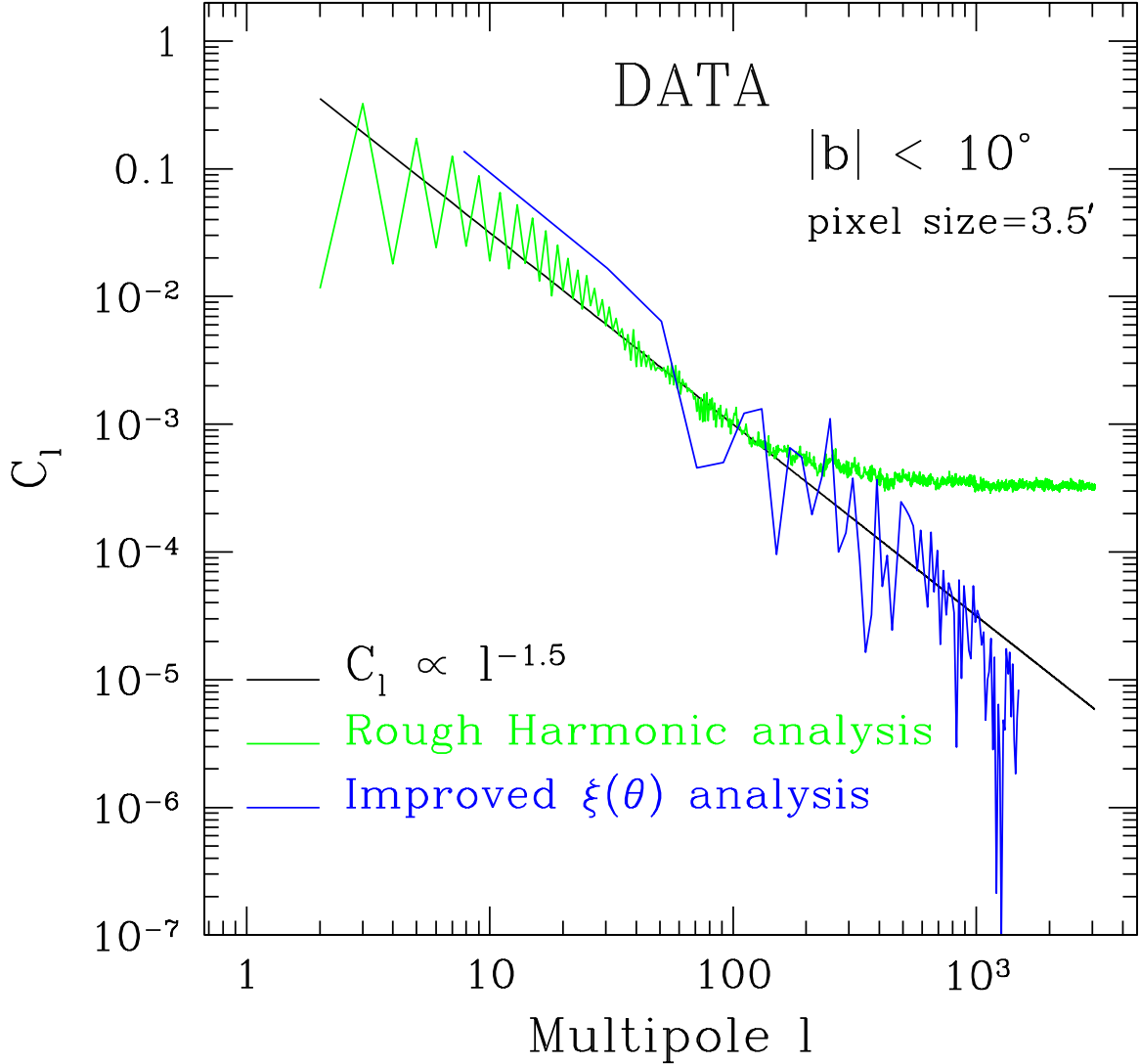}
\end{center}
\caption{\label{cl_star_data}
Angular power spectrum (PS) of the starlight polarization degree map
in the Galactic plane, $|{\rm b}| < 10^{\circ}$, which contains 4114 lines of sight.
The PS estimation from the rough harmonic analysis (green line) 
is dominated by shot-noise for multipoles $\ell \simgt 100$ i.e, 
angular scales 
$\theta \simlt 2^{\circ}$.  
The improved correlation function analysis (blue line), 
which corrects for shot-noise and pixel
window effects, shows that the underlying continuum signal can be well-fitted
by $C_{\ell} \propto \ell^{-1.5}$ (black line) 
down to multipoles $\ell \approx 1000$ i.e, 
scales $\theta \approx 10^{\prime}$.}
\end{figure*}

We aim at computing the PS of the 
starlight polarization degree data according to Eq(\ref{eq:cl}). 
In order to do so, we have generated 
a full-sky map (we use a HEALPix tessellation, see 
{\tt http://www.eso.org/kgorski/healpix) } 
for stars at $|{\rm b}| < 10^{\circ}$, which comprises 4114 stars, 
i.e, non-zero pixels (see Table \ref{tab:sample}).  
The resolution of the map has been chosen to be high enough 
($3.5^{\prime}$ pixels)
so that different sources are not identified with the same pixel.
The PS of this map, shown in Fig(\ref{cl_star_data})
(see green line), has been computed using 
the {\em anafast} program of the HEALPix package.
The slope of the PS is well-fitted by $C_{\ell} \propto \ell^{-1.5}$
down to $\ell \simlt 100$ which translates into angular scales 
$\theta \simgt 2^{\circ}$.
On smaller scales $\theta \simlt 2^{\circ}$, the PS is dominated 
by a flat shot-noise-like spectrum ($C_{\ell} = constant$), which
is due to the effect of the large number of quasi-randomly located zero-valued
pixels in the map, and which
prevents us from measuring the 
PS of the underlying continuum signal 
down to the pixel resolution scale. 

\subsubsection{Improved Analysis: the Correlation Function Approach} 
\label{sec:improved}

In order to improve the simple analysis presented above, we should
correct the data for the pixel window and the shot-noise intrinsic to
the sparsely distributed data we want to analyze.

For this purpose we first compute the two-point correlation function of the
polarization degree data using  
a quadratic estimator where the weighting is 
effectively done in pixel space \\ 
(Szapudi \& Szalay 1998; see also Szapudi et al 2001). 
The method consists of the following steps: 

\begin{itemize}

\item{
Compute the correlation function $\xi (\theta)$ of the
data on a very fine grid of $\theta$ (typically 300000 bins).}

\item{Resample the grid at
the roots of the Legendre polynomial of order $\ell_{max}$, where $\ell_{max}$ is
the maximum multipole at which one wishes to estimate the PS.
The re-sampling, using a Gaussian interpolating kernel on the fine bins, 
weighted by the number of pairs per bin, is at the core of the method
(see Szapudi et al 2001 for details).}

\item{Obtain the PS by
a simple Gauss-Legendre quadrature integration. The output is presented as 
flat band powers i.e, data averaged in  
multipole bins of width $\Delta \ell = 20$.} 

\end{itemize}

The advantage of this
method (with respect to the Harmonic approach) 
is that it gives an unbiased 
estimate of the power spectrum for an arbitrary sampling of the sky,
thus
avoiding the shot-noise power bias that is visible in the traditional PS 
estimator, as given by Eq(\ref{eq:cl}). 
In turn this typically translates into a noisy
estimate for the range of $\ell$'s where the shot-noise power dominates.

Fig \ref{cl_star_data} (see blue line) 
shows how this method allows for an efficient way of
de-convolving the window function\footnote{
This is effectively taken into account at a later stage, in multipole
space,
where $C_{\ell}$ is divided by the square of the pixel window
function} 
and 
removing the shot-noise component of the signal down to 
$\ell \approx 1000$, i.e, $\theta \approx 10^{\prime}$ which is close to the
map resolution $\theta \approx 3.5^{\prime}$. 
Although the estimated signal is rather noisy 
on small scales $\ell \simgt 100$, where the shot-noise dominates,
the PS can be well-fitted on average by a power-law $C_l \propto \ell^{-1.5}$
in the whole range of scales measured, $\ell \simlt 1000$.

\subsection{Power Spectrum Estimation of the Simulated Data}
\label{sec:sim}

We would like to know how robust are the results obtained 
in the previous section.
In particular, how sensitive is the estimated 
PS slope to the spatial distribution of stars used ?
how does the clustering of the sources affect the PS analysis ?
To answer these questions we 
have first simulated a {\em mock starlight map} in the Galactic plane
($|{\rm b}| < 10^{\circ}$) for which we have computed the PS, as follows:

\begin{itemize}

\item{Generate a random-Gaussian realization full-sky map 
of a $C_{\ell} \approx \ell^{-1.5}$ (with arbitrary amplitude), 
\footnote{We use the {\em synfast} program of the HEALPix package}
with the same spatial resolution than 
the original starlight data map ($3.5^{\prime}$ pixels). The latter PS 
gives a good fit to the starlight polarization degree PS for $\ell \simlt 1000$,
as shown in \S\ref{sec:pse}}.

\item{Set all pixels outside the Galactic plane 
$|{\rm b}| < 10^{\circ}$ to zero.}

\item{Remove at random all but 4114 of the Galactic plane ($3.5^{\prime}$) pixels
(or lines of sight) from the map. This leaves as many non-zero 
pixels from the random-Gaussian realization as 
in the original starlight data map (see Table \ref{tab:sample}).}

\item{Compute the PS of the resulting sparsely sampled 
{\em mock starlight map} for the polarization degree as we did for
the actual data according to the methods presented in \S\ref{sec:pse}.}

\end{itemize}

\begin{figure*}[t!]
\begin{center}
\includegraphics[width=16.0cm]{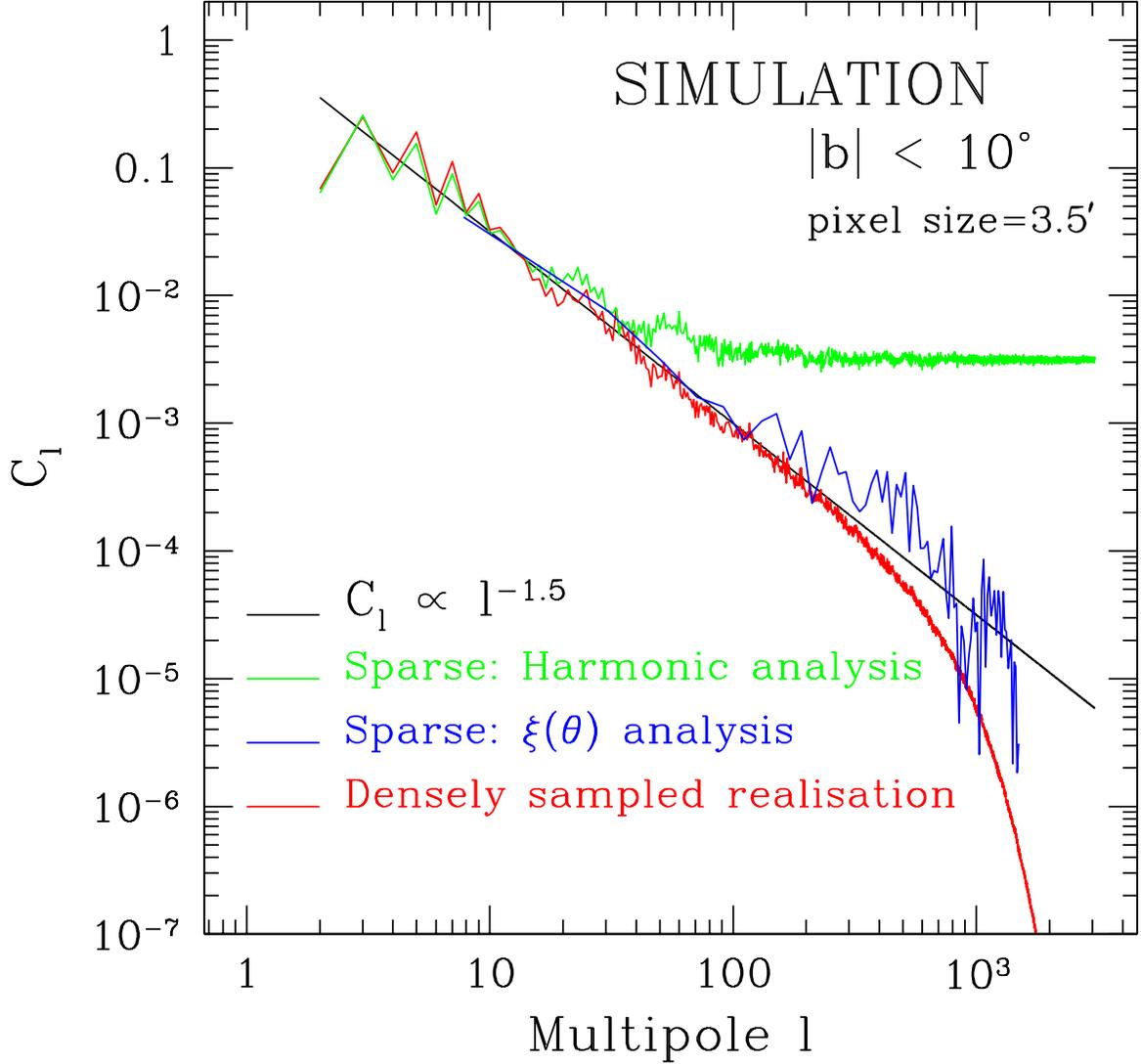}
\end{center}
\caption{\label{cl_star_sim}
Same as Fig \ref{cl_star_data} but for the {\em simulated} 
starlight polarization degree map.
The red line shows the densely-sampled (homogeneous) 
random-Gaussian realization
of $C_{\ell} \propto \ell^{-1.5}$. The damping tail for $\ell \simgt 200$ 
is due to the pixel window.
The PS estimation from the rough harmonic analysis (green line) 
is dominated by shot-noise for multipoles $\ell \simgt 40$ i.e, scales 
$\theta \simlt 5^{\circ}$, while the improved correlation function analysis
shows that the underlying continuum signal can be recovered
down to multipoles $\ell \approx 1000$ i.e, 
up to scales $\theta \approx 10^{\prime}$.}
\end{figure*}

As seen in Fig \ref{cl_star_sim} (see green line), the rough harmonic analysis 
(see \S\ref{sec:rough}) shows that the simulated starlight data has
approximately the same PS than the 
underlying densely sampled (effectively continuous) distributed data (red line)
for multipoles $\ell \simlt 40$ i.e, scales $\theta \simgt 5^{\circ}$.
On smaller scales, shot-noise dominates the signal.
However, the improved analysis based on the correlation function approach (see 
\S\ref{sec:improved}) allows to measure the PS of the underlying 
continuous distributed map
from the sparsely sampled {\em mock starlight map} down to much higher multipoles, 
$\ell \simlt 1000$ i.e, scales $\theta \simgt 10^{\prime}$ (see green line), 
although the signal is rather noisy beyond the shot-noise dominance scale
$\ell \sim 40$.

\subsection{Effect of Clustering}
\label{sec:clust}

Since the {\em mock starlight map} was generated from a random-Gaussian realization,
we can compare the PS analysis obtained from it (see \S\ref{sec:sim}) 
to the one performed on the real data (see \S\ref{sec:pse}), 
which are clearly non-Gaussian distributed (see Fig \ref{polDegree_5513}), 
to see what is the effect of clustering or 
non-Gaussianity in the distribution of the lines of sight.

Our findings in \S\ref{sec:sim} show that the clustering of the sources
in the analyzed starlight polarization degree map shifts the scale
where shot-noise dominates from $\theta \approx 5^{\circ}$ down to 
$\theta \approx 2^{\circ}$. Therefore, a rough harmonic analysis
allows a clean measurement
of the PS of the underlying continuously distributed map
down to much (a factor of 2.5) smaller scales when the sources are
strongly clustered. 
However, using the {\em improved} correlation
function analysis, shot-noise and pixel window effects can be effectively
removed down to approximately the map resolution scale ($3.5^{\prime}$)
with hardly any dependence on the clustering of the sources.
The price to pay for this is 
that the more clustered the sources are, the noisier
the measured PS turns out to be on small scales 
(see blue lines in Figs \S\ref{cl_star_data} \& \ref{cl_star_sim}).


\section{Discussion}
\label{sec:disc}

In this paper we present an statistical analysis of the largest compilation 
available of Galactic starlight polarization data.
The data analyzed consists of 5513 stars, a large fraction of which is found at
low Galactic latitudes 
($|{\rm b}| < 10^{\circ}$) and in the vicinity of the sun 
(d $<$ 1 \kpc).
Despite the inhomogeneous distribution of the sample, it allows for a proper 
statistical investigation of the large-scale behavior of the 
stellar parameters, such as the linear polarization (polarization degree and angle) 
and extinction, as induced by intervening dust grains. 

The correlations between stellar parameters give some valuable information 
on the geometry and degree of uniformity of the Galactic magnetic field, as we
discuss in \S\ref{sec:borient} \& \S\ref{sec:beff}.
On the other hand, we discuss to what extent starlight data 
traces the the large-scale pattern of the polarized ISM
in \S\ref{sec:diffuse}, based on our
PS analysis (\S\ref{sec:pse}).

\subsection{Polarization Angle and Magnetic Field Orientation}
\label{sec:borient}

Although no consensus has been reached 
in relation to what mechanism
is the dominant in the interstellar 
environment\footnote{Radiative torque mechanism looks the 
most promising right now (Draine \& Weingartner 1996, 
Draine \& Weingartner 1997).}
(see discussion in Lazarian 2000) it is generally accepted that 
grains in diffuse interstellar gas tend to be
aligned with their major axes perpendicular to the magnetic field. 
The cases where the alignment is suspected to be parallel
to magnetic field are extremely rare (see Rao et al 1998) and can
be safely ignored in our analysis.  

According to this picture, the electric field of radiation transmitted 
by an interstellar dust grain is less absorbed along
the grain minor axis and therefore polarized in that direction which is
parallel to the external magnetic field orientation.
Therefore, polarized starlight radiation vectors are oriented parallel
to the Galactic magnetic field.

Since starlight polarization vectors are only seen as projected 
in the plane of the sky, they just give us direct
information on the plane-of-the-sky projection of the 
Galactic magnetic field orientation.
In \S\ref{sec:londep} we found 
a strong alignment of starlight polarization vectors with the Galactic plane
structures and the spherical shell of Loop 1 (see \S\ref{sec:londep} \& 
\S\ref{sec:latdep}), which in turn provides evidence that {\em there is a 
net alignment of the magnetic field 
(as seen from its plane-of-the-sky projection) with 
Galactic structures on large-scales}.

However, we stress that 
the full reconstruction of the 3D magnetic field orientations (and strength) 
requires additional complementary data from radio (synchrotron), 
sub-mm/IR (dust) observations and rotational measures from distant pulsars 
(Zweibel \& Heiles 1997).

\subsection{Polarization Degree and Efficiency of Magnetic Field Alignment}
\label{sec:beff}

We have shown in our analysis
that the starlight polarization degree and extinction are clearly correlated and
on the mean, this correlation has a lower amplitude than 
what is expected from complete
dust-grain alignment from homogeneous magnetic fields (see \S\ref{sec:dist}).
The fact that starlight data exhibits a lower polarization degree 
as a function of extinction
than the theoretical upper limit, suggests that 
{\em either the grain alignment 
is not optimal or the Galactic magnetic field 
has a significant random component}.

For grains in clouds with high extinction the alignment tends to fail
(Lazarian et al 1997) but our sample does not deal with
such clouds. At the same time, measurements of high degree of polarization
suggest that the alignment can be sufficiently efficient. Substantial
variations of the grain properties (e.g, the degree of elongation)
do not look promising either.

On the other hand, a random component of the magnetic field ${\bf {B_r}}$ 
smears to some degree the correlation introduced by 
the uniform component ${\bf {B_u}}$
(Jones 1989). 
This smearing effect is likely to affect the observed stars
as supported by the high degree of incoherence (randomness)
observed for the starlight polarization angle (see Fig \ref{polAngle_5513}).
In what follows we shall assume that the latter is the dominant misalignment 
effect to infer an upper limit in
the degree of randomness of the Galactic magnetic field.

It is well-known that, by assuming a model, one 
can relate the observed polarization degree to the ratio of uniform to
random plane-of-the-sky components of the 
underlying magnetic field (see e.g, Heiles 1996).
In particular, assuming Burn's model (Burn 1966), 
one finds for the starlight sample, 
Eq(\ref{eq:polratio}),
\beq
{\rm {P_{Obs} \over P_{Max}}} = {{\bf B^2_u} \over {{\bf B^2_u} + 
{\bf B^2_r}}}  \approx 0.39-0.62 
\eeq
within the range, $0.1 \simlt$ E(B-V) $\simlt 1$, where the power-law is a good fit to 
${P_{Obs}/P_{Max}}$ indeed.
This yields, ${\bf {B_u}}/{\bf {B_r}} \approx 0.80-1.27$ for the same range.
Note that Burn's model
assumes a 3D distribution of the random component (consistent with 
a scenario with a uniform field distorted by supernovae explosions)
and a long enough path-length for every line of sight i.e, a large 
number of intervening clouds. 
In fact, for a subsample of nearby stars, the model
tends to overestimate the ${\bf {B_u}}/{\bf {B_r}}$ ratio (Heiles 1996).
Note that this bias is actually consistent with Eq(\ref{eq:polratio}) which
predicts a larger ${\rm {P_{Obs}/P_{Max}}}$ for low extinction (or distance)
and therefore, a larger  ${\bf {B_u}}/{\bf {B_r}}$ ratio.  

According to the above discussion, 
we take the value for high extinction E(B-V) $\simgt 1$, 
which corresponds to distant stars (d $>$ 1 \kpc, 
see lower panel of Fig \ref{disbins_poldeg_ebv}),
as an unbiased estimate of the ratio,  ${\bf {B_u}}/{\bf {B_r}} \approx 0.8$.
This value is roughly consistent with previous estimates from starlight data 
(see Heiles 1996 for a review and references therein):
${\bf {B_u}}/{\bf {B_r}} \approx 0.68$.
Note that ${\bf {B_u}}/{\bf {B_r}}$ as derived from starlight polarization data
is typically larger than 
estimates from synchrotron polarization, ${\bf {B_u}}/{\bf {B_r}} \approx 0.54$,
and it is significantly larger than that obtained from rotational measures
of distant pulsars, ${\bf {B_u}}/{\bf {B_r}} \approx 0.28$.

The discrepancy between estimates from different data sets
can be explained as every data set basically samples a different component
of the interstellar medium: 
pulsars mainly trace the warm ionized medium,
starlight data samples primarily the neutral media\footnote{
HI dominates the column density, although
H+ also contributes significantly (see Draine \& Lazarian 1998).}
while synchrotron data seems to sample all components  
(Heiles 1996).

\subsection{Large-scale Pattern of the Polarized ISM} 
\label{sec:diffuse}

We have performed an angular power spectrum (PS) analysis 
of the starlight polarization degree.
We have focused on Galactic plane data ($|{\rm b}| < 10^{\circ}$)
as it concentrates most of the sources in the sample and therefore makes 
the statistical analysis more reliable.
Our analysis shows that the starlight polarization degree PS 
is well fitted by a power-law behavior,  
$C_{\ell} \propto \ell^{-1.5}$ for $\ell \simlt 1000$ 
(where the multipole order 
${\ell} \approx 180^{\circ}/\theta$) which translates into 
angular scales ${\theta} \simgt 10^{\prime}$.
This is approximately the pixel resolution scale used, 
$3.5^{\prime}$ (see \S\ref{sec:pse}). 
This result was obtained thanks to a correlation function method
which efficiently removes the shot-noise and pixel window effects
that strongly affect the data on small scales (see \S\ref{sec:improved}).

We have assessed how the above results are affected by the 
{\em clustering} or non-Gaussianity in the distribution of sources by
simulating a {\em mock starlight map} for the polarization degree 
(see \S\ref{sec:clust}).
We found that the efficiency with which one measures the PS of the 
underlying densely-sampled signal is not significantly altered by 
the clustering of the sources, although for the 
real non-Gaussian distributed sources
shot-noise dominates at smaller scales and the estimated PS is noisier
than the simulated random-Gaussian case.

The above results provide evidence that the use
of the polarization degree of the ISM as 
sparsely-sampled from 
lines-of-sight to several thousand (galactic-plane) stars 
allows a clean reconstruction of the PS
of an underlying {\em homogenelously sampled} (continuum) 
polarization degree of the ISM.
In particular, we find that the ISM 
polarization degree in the continumm has the same PS slope
than that measured from sparsely-sampled data,  
$C_{\ell} \propto \ell^{-1.5}$.

The PS of the ISM dust polarization pattern 
happens to have a similar slope than that
of the synchrotron Galactic polarization. Indeed, Baccigalupi et al
(2001) showed that in the range of $1000>\ell>100$ the spectrum of polarised
synchrotron continuum scales as $\ell^{1.7}$ with the uncertainty in
the index $\pm 0.3$, which is compatible with the PS slope we find for the
starlight polarization degree. 
However, synchrotron and aligned dust trace different phases of
the ISM and the emission has, in principle, a completely different nature.
Thus, on the first glance, one does not expect to see any
correspondence between the two spectra. But if {\em variations} 
in the polarization arise mainly because of variations in the Galactic
magnetic field, the properties of the two may be related. 
Further study shall show whether this similar statistical behaviour 
can be explained on physical grounds.

On the other hand,
the lack of starlight data on high-galactic latitudes does not allow 
to make a reliable measurement of the PS for the moment. 
In particular, it remains to be seen 
whether the PS slope varies significantly at high-galactic latitudes
in the same way as has been found in recent analyses from synchrotron emission
(Baccigalupi et al 2000).


In a forthcoming paper (Fosalba et al 2001), 
we shall use our results on starlight data to
model Galactic polarized dust emission at sub-mm/FIR wavelengths 
and its effect on the process of foreground subtraction in 
cosmic microwave background experiments.

\acknowledgments
We acknowledge the use of the starlight data compilation by C. Heiles
who has kindly made it publicly available. 
PF is supported by a CMBNET fellowship of the European Comission.

\cleardoublepage



\begin{deluxetable}{cccccc}
\tablecaption{Mean Stellar Parameters: Sample of 5513 Stars\label{tab:sample}}
\tablehead{
              \colhead{Latitude}
             &\colhead{Distance}
             &\colhead{Stars $(\%)$}
             &\colhead{P($\%$)}
             &\colhead{E(B-V)} 
             &\colhead{$\theta_p$}
}
\startdata
      			 & Total & $4114 (75)$ & $1.69$  & $0.49$ & $88.1$ \\
Low Latitude		 & Nearby & $1451 (26)$ & $0.94$ & $0.29$ & $90.4$ \\
      			& Distant & $2663 (48)$ & $2.09$ & $0.60$ & $86.9$ \\
\tableline
  			& Total  & $1399 (25)$ & $0.45$ & $0.15$ & $93.0$ \\
High Latitude	        & Nearby & $1315 (24)$ & $0.42$ & $0.14$ & $93.4$ \\
              		& Distant & $84 (1)$ & $0.89$ & $0.26$ & $86.7$ \\
\enddata
\tablecomments{Mean stellar parameters i.e, polarization degree P($\%$), 
extinction E(B-V), and polarization
angle $\theta_p$, for the sample of 5513 stars analyzed,
as a function of latitude and distance. High latitude (low latitude)
means $|{\rm b}| > 10^{\circ}$ ($|{\rm b}| < 10^{\circ}$) and nearby (distant) denotes d $<$ 1 \kpc ~(d $>$ 1 \kpc).}
\end{deluxetable}

\cleardoublepage


\cleardoublepage


\end{document}